\documentclass[pra,unsortedaddress,superscriptaddress,a4paper,nofootinbib,%
nobalancelastpage,preprintnumbers]{revtex4}%
\usepackage{epsfig}
\usepackage{graphicx}
\usepackage{graphics}
\usepackage{xspace}
\usepackage{amssymb}
\usepackage{amsmath}
\usepackage{latexsym}
\usepackage{natbib}
\usepackage{mathrsfs}
\usepackage{url}

\def\q{\mbox{\boldmath $q$}}
\def\p{\mbox{\boldmath $p$}}
\def\r{\mbox{\boldmath $r$}}
\def\J{\mbox{\boldmath $J$}}

\begin{document}

\title{Electromagnetic proton-neutron knockout off $^{16}$O: new achievements
 in theory}

\author{ Carlotta Giusti}
\affiliation{Dipartimento di Fisica Nucleare e Teorica,
Universit\`a degli Studi di Pavia\\ 
Istituto Nazionale di Fisica Nucleare,
Sezione di Pavia, I-27100 Pavia, Italy}
\author{ Franco Davide Pacati}
\affiliation{Dipartimento di Fisica Nucleare e Teorica,
Universit\`a degli Studi di Pavia\\ 
Istituto Nazionale di Fisica Nucleare,
Sezione di Pavia, I-27100 Pavia, Italy}
\author{Michael Schwamb}
\affiliation{ Dipartimento di Fisica, Universit\`{a} degli Studi di Trento \\
and Istituto Nazionale di Fisica Nucleare,
Gruppo Collegato di Trento,  I-38100 Povo (Trento), Italy }
\affiliation{ European Center for Theoretical Studies in Nuclear
Physics and Related Areas (ECT$^{\ast}$),
I-38100 Villazzano (Trento), Italy}
\author{Sigfrido Boffi}
\affiliation{Dipartimento di Fisica Nucleare e Teorica,
Universit\`a degli Studi di Pavia\\ 
Istituto Nazionale di Fisica Nucleare,
Sezione di Pavia, I-27100 Pavia, Italy}

\begin{abstract}
Results for the cross sections of 
the exclusive $^{16}$O(e,e$'$pn)$^{14}$N and 
$^{16}$O($\gamma$,pn)$^{14}$N knockout reactions are presented and discussed 
in different kinematics.  In comparison with earlier work, a complete treatment 
of the center-of-mass (CM) effects in the nuclear one-body current is 
considered in connection with the problem of the lack of orthogonality between 
initial bound and final scattering states. The effects due to CM and
orthogonalization are investigated in combination with different treatments of
correlations in the two-nucleon overlap function and for different
parametrizations of the two-body currents. The CM effects lead in 
super-parallel kinematics to a dramatic increase of the $^{16}$O(e,e$'$pn) 
cross section to the $1_2^+$ excited state (3.95 MeV) of $^{14}$N. In all the 
situations considered the results are very sensitive to the treatment of 
correlations. A crucial role is played by tensor correlations, but also the
contribution of  long-range correlations is important. 

\end{abstract}

\maketitle

\section{Introduction}
\label{sec:intro}
It is well known that the independent particle shell model, in which nucleons
 move in a mean field, is not sufficient to describe all basic properties
 of atomic  nuclei. This failure is a consequence
 of the strong short-range component of the NN-interaction, which induces
 into the nuclear wave function components beyond the mean field description.
 Thus, a careful evaluation of the corresponding short-range correlations
 (SRC) is inevitable for the understanding of 
 nuclear structure in general \cite{MuP00}. 
 
Electromagnetically induced two-nucleon knockout is apparently the most 
promising   tool to study SRC \cite{Ox}. However, competing two-body mechanisms
like meson-exchange currents (MEC), isobar-excitation or final state 
interactions (FSI) need to be taken into account as well in order to obtain a 
realistic description of the process. 
 These ``background'' effects 
 must be well under control in order to extract from experiment any information
 about the fundamental correlations. 
This requires a theoretical approach which should be as comprehensive as  
possible.  Presently, different models are available
(see \cite{AnC04,RyN04,sch2} and references therein).

In the past years, the Pavia group has improved, step by step,
 its model for two-nucleon knockout. Recent improvements have been performed
 with respect to the treatment of FSI \cite{sch1a,sch1}, of the two-nucleon
 overlap  function \cite{barb}, of the $\Delta$-current contribution 
 \cite{sch2}, and, finally, of the center-of mass (CM) effects in the 
 electromagnetic current operator \cite{cm}. With respect to the last two
 issues, only  proton-proton (pp) knockout has been considered,
 which is conceptually simpler than proton-neutron (pn)
 knockout due to the suppression of MEC contributions.
 However, pn-knockout is of specific interest for the 
 study of tensor correlations (TC), which are due to the strong tensor 
 component of the pion-exchange contribution to the NN-interaction.

 In the present paper, we study thoroughly  the impact
 of the above mentioned conceptual improvements on pn-knockout.
 The final aim -- in the long term run -- is to develop an approach which is
 as comprehensive as practically possible, so that hopefully important
  conclusions about
 the structure of correlations can be drawn by comparison with experiment. 
 
The paper is organized as follows. In the next section, the main features
 of the theoretical approach  are outlined. Special emphasis
  is devoted to the description of the various improvements performed 
  in comparison with earlier work.
Numerical results for the exclusive $^{16}$O(e,e$'$pn)$^{14}$N and 
$^{16}$O($\gamma$,pn)$^{14}$N reactions in different kinematics are 
presented in sect.~\ref{sec:results}. Some conclusions are drawn in 
 sect.~\ref{sec:conclusions}.
 
 Finally, we want to stress the purely theoretical nature of the paper. 
 A comparison with recent experimental data \cite {middleton} is presently 
 under investigation and will be discussed in forthcoming work \cite{secondpn}.

\section{Theoretical model}
\label{sec:reaction}
The cross section of a reaction induced by a real or virtual photon, with momentum 
$\q$, where two nucleons, with momenta $\p'_{1}$, and $\p'_{2}$, are ejected 
from a nucleus, can be written in terms of the transition matrix elements of 
the charge-current density operator between initial and final nuclear states 
\begin{equation}
J^\mu (\q) = \int  
< \Psi_{\mathrm{f}} | \hat{J}^\mu(\r) |
\Psi_{\mathrm{i}} >
{\mathrm{e}}^{\,{\mathrm{i}}{\footnotesize \q} \cdot
{\footnotesize \r}} {\mathrm d}\r.
\label{eq:jm}
\end{equation} 
Bilinear products of these integrals give the components of the hadron tensor  
and therefore the cross section \cite{Ox,GP}.

For an exclusive process, where the residual nucleus is left in a discrete
eigenstate of its Hamiltonian, and under the assumption of a direct knock-out 
mechanism, the matrix elements of eq.~(\ref{eq:jm})
can be written as~\cite{GP,GP97}\footnote{Spin/isospin indices are generally 
suppressed in the formulas of this paper for the sake of simplicity.}
\begin{eqnarray}
& J^{\mu}({\mbox{\boldmath $q$}}) & = \int
{ \psi}_{\rm{f}}^{*}
({\mbox{\boldmath $r$}}_{1},{\mbox{\boldmath $r$}}_{2})
J^{\mu}({\mbox{\boldmath $r$}},{\mbox{\boldmath $r$}}_{1},
{\mbox{\boldmath $r$}}_{2})
\nonumber \\
 & & \times \, {\psi}_{\rm{i}}
({\mbox{\boldmath $r$}}_{1},{\mbox{\boldmath $r$}}_{2})
{\rm{e}}^{{\rm{i}}
\hbox{\footnotesize {\mbox{\boldmath $q$}}}
\cdot
\hbox{\footnotesize {\mbox{\boldmath $r$}}}
} {\rm d}{\mbox{\boldmath $r$}} \,
{\rm d}{\mbox{\boldmath $r$}}_{1} {\rm d}{\mbox{\boldmath $r$}}_{2}. 
 \label{eq:jq}
\end{eqnarray}

Three main ingredients appear in eq.~(\ref{eq:jq}):  the 
two-nucleon overlap integral ${\psi}_{\rm{i}}$ 
between the ground state of the target and the final state of the residual 
nucleus, the nuclear current $J^{\mu}$, and the two-nucleon scattering 
wave function ${\psi}_{\rm{f}}$.

In principle, the bound and scattering states, $\psi_{\rm{i}}$ and 
$\psi_{\rm{f}}$, are consistently derived from an energy-dependent non-Hermitian 
Feshbach-type Hamiltonian for the considered final state of the residual 
nucleus. They are eigenfunctions of this Hamiltonian at negative and positive 
energy values \cite{Ox,GP}.  
In practice, it is not possible to achieve this consistency and the
treatment of initial and final states proceeds separately with different 
approximations.

The two-nucleon overlap function (TOF) $\psi_{\rm{i}}$ contains information on 
nuclear structure and correlations. For the case of proton-neutron emission   
from $^{16}$O different approaches are used in \cite{barb,GMPS,GPpn} to
calculate the TOF. 

In the more sophisticated approach of \cite{barb}, that we call SF-B, the TOF 
was computed partitioning the Hilbert space in order to determine the 
contribution of long-range correlations (LRC) and SRC separately. The LRC, 
describing the  collective motion at low energy as well as the long range part 
of TC, were computed using the self-consistent Green's function formalism 
\cite{BD} in an appropriate harmonic-oscillator (h.o.) basis. The effects of 
SRC due to the central and tensor part at high momenta were added by computing 
the appropriate defect functions with the Bonn-C NN-potential \cite{MaH87}.

In the approach of \cite{GMPS}, that we call SF-CC, the effects of SRC as well 
as TC were calculated within the framework of the coupled cluster method, 
using the so-called $S_2$ approximation and employing the Argonne V14 
potential \cite{v14} for the NN-interaction. In this calculation the effects 
of LRC are accounted for in a simpler way and only knockout of nucleons from 
the $0p$  shell is considered. 

A much simpler approximation is used in \cite{GPpn}, where the TOF is given 
by the product of a coupled and antisymmetrized shell model pair function and  
of a Jastrow-type central and state independent correlation function taken 
from \cite{GD}. In this approach, that we call SM-SRC, only SRC are considered 
and the final state of the residual nucleus is a pure two-hole state 
in $^{16}$O.

The final-state wave function $\psi_{\rm{f}}$ is written as the product of two 
uncoupled s.p. distorted wave functions, eigenfunctions of a complex 
phenomenological optical potential which contains a central, a Coulomb, and a 
spin-orbit term \cite{Nad81}.
The effect of the mutual interaction between the two outgoing nucleons (NN-FSI) 
has been studied in \cite{sch1a,sch1,KM} and is included in the present  
calculations only in some cases using the model applied in \cite{sch1}. 

The nuclear current $J^{\mu}$ is the sum of a one-body  (OB) and a two-body (TB) 
contribution, {\it i.e.}
\begin{eqnarray}
J^{\mu}({\mbox{\boldmath $r$}},{\mbox{\boldmath $r$}}_{1},
{\mbox{\boldmath $r$}}_{2}) &=& J^{(1)\mu}({\mbox{\boldmath $r$}},
{\mbox{\boldmath $r$}}_{1}) + J^{(1)\mu}({\mbox{\boldmath $r$}},
{\mbox{\boldmath $r$}}_{2}) \nonumber \\
&+& J^{(2)\mu}({\mbox{\boldmath $r$}},
{\mbox{\boldmath $r$}}_{1},{\mbox{\boldmath $r$}}_{2}).
\label{eq:jq12}
\end{eqnarray}
The OB part includes the longitudinal charge term and the transverse 
convective and spin currents.
The TB current is derived from the effective Lagrangian 
of \cite{Peccei}, performing a non relativistic reduction of the lowest-order 
Feynman diagrams with one-pion exchange. We have thus currents corresponding 
to the seagull and pion-in-flight diagrams, and to the diagrams with 
intermediate $\Delta$-isobar configurations~\cite{gnn}, {\it i.e.}
\begin{eqnarray}
& &\J^{(2)}(\r,\r_{1},\r_{2})  = 
\J^{\mathrm{sea}}(\r,\r_{1},\r_{2}) \nonumber \\
& & \ + \ \J^{\pi}(\r,\r_{1},\r_{2}) 
 +  \J^{\Delta}(\r,\r_{1},\r_{2}) . \label{eq:nc}
\end{eqnarray}

Details of the nuclear current components can be found in 
\cite{sch2,gnn,WiA97,mec}. In this paper, the parameters of the 
$\Delta$-current are  fixed considering the NN-scattering in the 
$\Delta$-region, where a fairly good 
description of the scattering data can be achieved by choosing  parameters 
similar to the ones of the full Bonn potential \cite{MaH87,Wil92}. This choice 
gives the parametrization ``$\Delta$(NN)'' discussed in  \cite{sch2} where 
 both $\pi$- and $\rho$-exchange are considered. 
 It turns out that a comparable description of the NN-scattering data can be 
 achieved in a conceptually simpler manner considering only $\pi$-ex\-cha\-nge. 
 In this choice, which is adopted in the present paper, the coupling constants
$f^2_{\pi NN}/( 4\pi) =0.078$ and $f^2_{\pi N \Delta}/( 4\pi) =0.35$
 are used with a dipole form factor ($n_{\pi NN}$=$n_{\pi N \Delta}$=1) 
using the  cutoffs
 $\Lambda_{\pi NN}= \Lambda_{\pi N \Delta}=700$ MeV.\footnote{The notation
  for the coupling constants and cutoffs is the same as in \cite{sch2}.} 
 With respect to the nonresonant pion-in-flight and seagull MEC, we use a 
 dipole cutoff  of 3 GeV in accordance with the Bonn-C potential
 \cite{MaH87}. This large value leads to an upper estimate of the role
 of these contributions.   In the following, 
 we denote this choice of parameters for the TB currents as the 
 ``Bonn parametrization''. 

In the previous calculations of \cite{barb,GMPS,GPpn} the same coupling 
constants were used with a simpler regularization in coordinate space, both 
for the MEC and the $\Delta$-current, which in practice is similar to 
the unregularized prescription for the $\Delta$-current  ``$\Delta$(NoReg)'' 
of \cite{sch2}.
Although a bit simplistic, we denote in the following this previous
prescription in  shorthand  as ``unregularized  parametrization''. 
The results obtained with the two parametrizations are compared in the next 
section.

In order to evaluate the transition amplitude of eq.~(\ref{eq:jq}), for the 
three-body system consisting of the two nucleons, 1 and 2, and of the residual 
nucleus $B$, it appears to be natural to work with CM coordinates 
\cite{cm,GP,jack} 
\begin{eqnarray}
&{\mbox{\boldmath $r$}}_{1B}& = {\mbox{\boldmath $r$}}_{1} - 
{\mbox{\boldmath $r$}}_{B}, \,\,{\mbox{\boldmath $r$}}_{2B} = 
{\mbox{\boldmath $r$}}_{2} - {\mbox{\boldmath $r$}}_{B}, \nonumber \\
&{\mbox{\boldmath $r$}}_{B}& = \sum _{i=3} ^{A} {\mbox{\boldmath $r$}}_{i}
/ (A-2).
 \label{eq:cm}
\end{eqnarray}
The conjugated momenta are given by
\begin{eqnarray}
{\mbox{\boldmath $p$}}_{1B} &=& \frac{A-1}{A}
{\mbox{\boldmath $p'$}}_{1} - \frac{1}{A} {\mbox{\boldmath $p'$}}_{2}
- \frac{1}{A} {\mbox{\boldmath $p$}}_{B} \,\, , \\
{\mbox{\boldmath $p$}}_{2B} &=& -\frac{1}{A}
{\mbox{\boldmath $p'$}}_{1} + \frac{A-1}{A} {\mbox{\boldmath $p'$}}_{2}
- \frac{1}{A} {\mbox{\boldmath $p$}}_{B} \,\, , \\
{\mbox{\boldmath $P$}} &=& 
{\mbox{\boldmath $p'$}}_{1} + {\mbox{\boldmath $p'$}}_{2}
+  {\mbox{\boldmath $p$}}_{B} \,\, , \label{eq:mom}
\end{eqnarray}
where  $\p_{B}=\q-\p'_1-\p'_2$ is the momentum of the residual nucleus in the
laboratory frame.

Applying this transformation to the matrix element of eq.~(\ref{eq:jq}) is
essentially equivalent \cite{cm} to substitute in the equation the operator 
$\exp \left( {\mathrm{i}} {\mbox{\boldmath $q$}}\cdot {\mbox{\boldmath $r$}}_{1}\right)$ 
with
\begin{eqnarray}
\exp \left(  {\mathrm{i}} {\mbox{\boldmath $q$}} \cdot 
\frac {A-1}{A}{\mbox{\boldmath $r$}}_{1B}
\right) \exp \left(- {\mathrm{i}} {\mbox{\boldmath $q$}} \cdot \frac {1}{A}
{\mbox{\boldmath $r$}}_{2B}\right).
\label{eq:opcm}
\end{eqnarray}
An analogous substitution applies to 
$\exp \left( {\mathrm{i}} {\mbox{\boldmath $q$}}\cdot 
{\mbox{\boldmath $r$}}_{2}\right)$ \cite{cm}.

In spite of the fact that a OB operator cannot affect two particles if they are 
not correlated, it can be seen from eq.~(\ref{eq:opcm}) that in the CM
frame the transition operator becomes a two-body operator even in the case of a 
OB nuclear current. 
Only in the limit $A \rightarrow \infty$ CM effects are neglected and the 
expression in eq.~(\ref{eq:jq}) vanishes for a pure OB current, when the matrix
element is calculated using orthogonalized single particle (s.p.) wave 
functions.
This means that, due to this CM effect, for finite nuclei the OB current can 
give a contribution to the cross section of two-particle emission independently 
of correlations. 
This  effect is similar to the one of the effective charges in electromagnetic 
reactions~\cite{ec}.

The matrix element of eq.~(\ref{eq:jq}) contains a spurious 
contribution since it does not vanish when the transition operator is set 
equal to 1. This is due to the lack of orthogonality between the initial and
the final state wave functions. The use of an effective nuclear
current operator would remove the orthogonality defect besides taking into 
account space truncation effects \cite{Ox,ort}. 
In the usual approach, however, the 
effective operator is replaced by the bare nuclear current operator.  
Thus, it is this replacement which may introduce a
spurious contribution which is not specifically due to the different 
prescriptions adopted in practical calculations, but is already present in  
eq.~(\ref{eq:jq}), when $\psi_{\rm{i}}$ and $\psi_{\rm{f}}$ are 
eigenfunctions of an energy-dependent Feshbach-type Hamiltonian corresponding 
to different energies. 

This spuriosity can be removed by subtracting from the transition amplitude 
the contribution of the OB current without correlations in the nuclear wave 
functions. This prescription was adopted in \cite{barb,GMPS,GPpn}. 
In this way, however, not only the spuriosity is subtracted, but also the CM 
effect given by the two-body operator in eq.~(\ref{eq:opcm}), which is present 
in the OB current independently of correlations and which is not spurious. 
The relevance of this effect was investigated in \cite{cm} for the case of
pp-knockout.

A more accurate procedure to get rid of the spuriosity is to 
enforce orthogonality between the initial and final states by means of a 
Gram-Schmidt orthogonalization \cite{ortho}. In this approach each one of 
the two s.p. distorted wave functions is orthogonalized to all the s.p. 
shell-model wave functions that are used to calculate the TOF, {\it i.e.}, for 
the TOF of \cite{barb}, to the h.o. states of the basis used in the calculation 
of the spectral function, which range from  the $0s$ up to the $1p$-$0f$ shell.
This procedure allows us to get rid of the spurious contribution to two-nucleon 
emission due to a OB operator acting on either nucleon of an 
uncorrelated pair, which is due to the lack of orthogonality between the s.p. 
bound and scattering states of the pair. In this approach, in consequence,
 no OB current contribution without correlations needs to be subtracted.
Moreover, it allows us to include automatically all CM effects via 
eq.~(\ref{eq:opcm}). This procedure was proposed and applied in \cite{cm} to
electromagnetic pp-knockout and is applied in this paper to 
electromagnetic pn-knockout reactions.

\section{Results}
\label{sec:results}

In this section, numerical results are presented for the cross sections of the  
$^{16}$O(e,e$'$pn)$^{14}$N and $^{16}$O($\gamma$,pn)$^{14}$N reactions
to discrete low-lying states in the residual nucleus.
The main aim of this study is to investigate the effects of CM and 
orthogonalization in combination with different treatments of correlations in
the TOF and for different parametrizations of the TB currents. The effects
of the mutual interaction between the two outgoing nucleons is also 
considered with a few examples.

Calculations have been performed in different situations and kinematics. 
Of particular interest for our study is the case of the 
super-parallel kinematics, where for two-proton knockout the CM effects due to 
the OB current without correlations are particularly large \cite{cm}. 

In the so-called super-parallel kinematics the two nucleons are ejected 
parallel and anti-parallel to the momentum transfer and, for a fixed value of 
the energy transfer $\omega$ and of the momentum transfer $q$,  it is possible to 
explore, for different values of the kinetic energies of the outgoing nucleons, 
all possible values of the recoil or missing momentum $p_B$. 
This kinematical setting, which is particularly favourable to emphasize  
short-range effects, has been widely investigated in our previous work, both 
for pp- and pn-knockout 
\cite{sch2,sch1a,sch1,barb,GP,GP97,GMPS,GPA98,Kadrev}, and is very interesting 
from the experimental point of view. In fact, the recent  
$^{16}$O(e,e$'$pp)$^{14}$C  \cite{Rosner} and  $^{16}$O(e,e$'$pn)$^{14}$N 
 \cite{middleton} experiments carried out at MAMI were both centred on the 
same super-parallel kinematics.
In  \cite{middleton} the data of the first measurements of the exclusive 
$^{16}$O(e,e$'$pn)$^{14}$N reactions are not described by the theoretical 
model of \cite{barb}. Large discrepancies, both in shape and magnitude, are 
found between theoretical and experimental cross sections at low missing 
momenta. 
It is therefore of particular interest to investigate the relevance of the CM 
effects included in the present approach in comparison with the results of
 \cite{barb}.
 
\begin{figure}
\centerline{
\resizebox{0.55\textwidth}{!}{
  \includegraphics{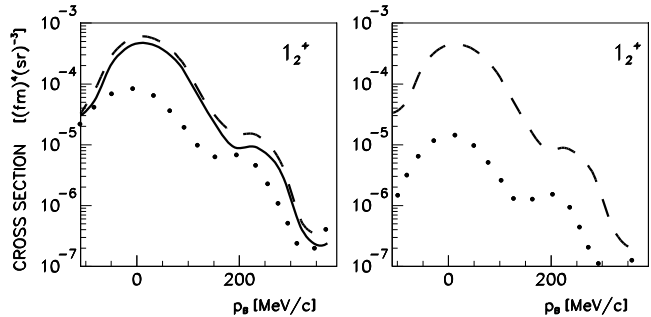}}
}
\caption{
The differential cross section of the $^{16}$O(e,e$'$pn)$^{14}$N reaction to 
the $1_2^+$ excited state (3.95 MeV) of $^{14}$N as a function of $p_B$ in a 
super-parallel kinematics with $E_0= 855$ MeV, electron scattering angle 
$\theta_e = 18^{\circ}$, $\omega=215$ MeV, and  $q=316$ MeV/$c$. The proton is
emitted parallel and the neutron antiparallel to the momentum transfer.
Different values of $p_B$ are obtained changing the kinetic energies of 
the outgoing nucleons. Positive (negative) values of $p_B$ refer to 
situations where $\p_B$ is parallel (anti-parallel) to $\q$. 
The final results given by the sum of the OB and the TB currents 
are displayed in the left panel, the separate contribution of the OB  
current is shown in the right panel. The TOF from the two-nucleon
spectral function of \cite{barb} (SF-B) is used in the calculations.
The dotted lines give the results of \cite{barb}, the dashed and 
solid lines are obtained with  the present approach, where the  
orthogonalization of s.p. bound and scattering states is enforced and all  
CM effects are taken into account. The solid and dashed lines in the left 
panel are obtained  with different parametrizations of the TB currents, 
{\it i.e.} our previous unregularized pararametrization, as in \cite{barb}, 
(dashed) and the Bonn parametrization (solid).
}
\label{pncm1}
\end{figure}

The comparison is shown in fig.~\ref{pncm1} for the cross section of the 
$^{16}$O(e,e$'$pn)$^{14}$N reaction to the $1_2^+$ excited state of $^{14}$N 
at 3.95 MeV. This is the state that is mostly populated in pn-knockout 
\cite{middleton,Isak,Garrow}. Results for this transition are 
therefore of particular interest. Calculations have been performed 
in the same super-parallel kinematics already considered in \cite{barb} 
and realized in the experiment \cite{middleton} at MAMI, {\it i.e.} the incident
electron energy is $E_{0}=855$ MeV, $\omega=215$ MeV, and $q=316$ MeV/$c$.
The CM contribution included in the orthogonalized approach produces a huge 
enhancement of the cross section calculated with the OB current.  The results 
are shown in the right panel of the figure. The enhancement is larger than one
order of magnitude at low missing momenta and is only slightly reduced at 
higher momenta. The difference between the cross section obtained in the
orthogonalized approach and the one of \cite{barb} is reduced when also the TB 
currents are included in the calculations. The results are shown in the left 
panel of the figure. In the approach of \cite{barb} the cross section is 
dominated by the TB currents, in particular by the $\Delta$-current. In
contrast, the OB current dominates the cross section in the orthogonalized 
approach and only a small enhancement is given by the TB currents. 
The differences obtained in the orthogonalized approach with the two 
prescriptions for the TB currents are appreciable although not very important. 
The Bonn parametrization reduces the cross section by at most 30-40 \%.  
In the final cross section the difference between the results of the two
approaches remains large, a bit less than one order of magnitude at low 
momenta, in the maximum region, and it is still sizable, although considerably 
reduced, at higher momenta.

\begin{figure}
\centerline{
\resizebox{0.55\textwidth}{!}{
  \includegraphics{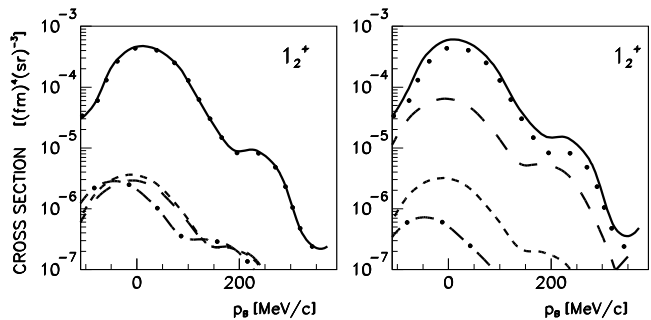}}
}
\caption{
The differential cross section of the $^{16}$O(e,e$'$pn)$^{14}$N reaction  
to the $1_2^+$ state as a function of $p_B$ in the same super-parallel 
kinematics as in fig.~\ref{pncm1}. Calculations are performed in the
orthogonalized approach. TOF as in fig.~\ref{pncm1}. Separate
contributions of the OB, seagull, pion-in-flight and $\Delta$-current 
are shown by the dotted, short-dashed, dotted-dashed, and long-dashed lines, 
respectively. The solid lines give the total cross section. The Bonn
parametrization  is used for the TB currents in the left panel, the
unregularized parametrization in the right panel.
}
\label{pncm2}
\end{figure}

The results given by the two parametrizations on the separate contributions of
the different components of the TB current are displayed in fig.~\ref{pncm2} 
and compared with the contribution of the OB current and with the final cross
section. The calculations have been performed with the orthogonalized approach 
for the $^{16}$O(e,e$'$pn)$^{14}$N reaction to the $1_2^+$ state in the 
super-parallel kinematics and with the SF-B overlap function. 
The CM contribution included in the orthogonalized approach gives large effects 
on the OB current and, as it has been shown in fig.~\ref{pncm1}, on the final
cross section. In contrast, the use of orthogonalized s.p. bound and scattering
wave functions produces only negligible effects on all the terms of the TB 
current. Therefore the results for the separate TB components obtained with the
unregularized 
 parametrization used in \cite{barb}, and displayed in the right panel of 
fig.~\ref{pncm2}, are practically the same as in \cite{barb}. The differences
obtained with the Bonn parametrization, which are displayed in the left panel 
of the figure, are appreciable for the seagull and pion-in-flight MEC and huge 
  for the $\Delta$-current, whose contribution is dramatically reduced with 
  the Bonn parametrization. 
In both cases, however, the cross section is dominated by the OB current.
With the Bonn parametrization the role of the TB currents is negligible and the
final cross section is in practice entirely due to the OB current. With the
parametrization used in the right panel of the figure the contribution of the
$\Delta$-current enhances the OB cross section by 30-40\% and is responsible 
for the difference between the two final results.

\begin{figure}
\centerline{
\resizebox{0.55\textwidth}{!}{
  \includegraphics{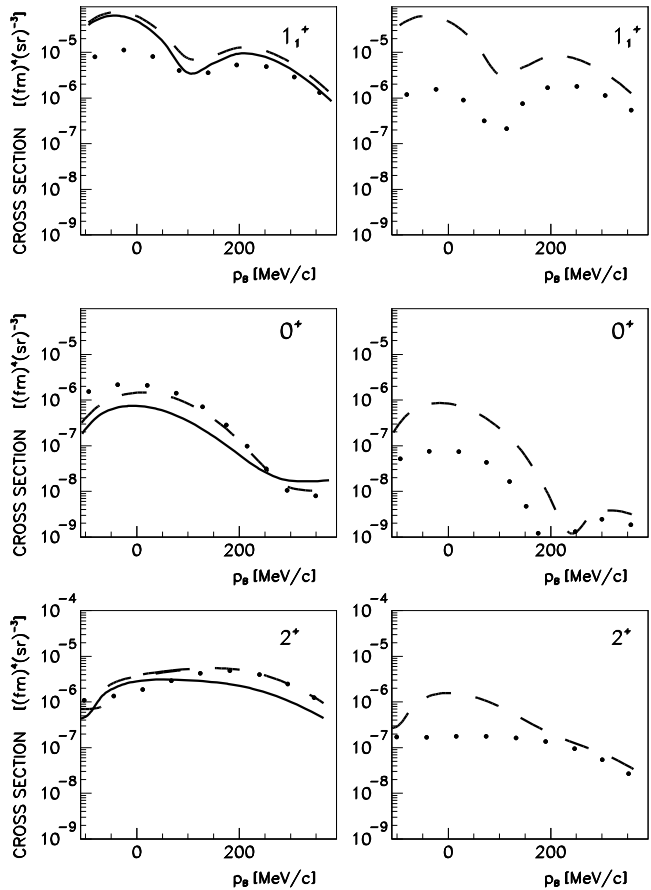}}
}
\caption{
The differential cross section of the 
$^{16}$O(e,e$'$pn)$^{14}$N  reaction  
to the $1_1^+$ ground state, the $0^+$ (2.31 MeV), and the $2^+$ (7.03 MeV)
excited states of $^{14}$N as a function of $p_B$ in the same super-parallel 
kinematics as in fig.~\ref{pncm1}. TOF as in fig.~\ref{pncm1}. Line convention
for left and right panels as in fig.~\ref{pncm1}.
}
\label{pncm3}
\end{figure}
In  fig.~\ref{pncm3} the cross sections of the $^{16}$O(e,e$'$pn)$^{14}$N  
reaction are displayed for transitions to different final states of the 
residual nucleus: the $1_1^+$ ground state, the $0^+$ (2.31 MeV), and the 
$2^+$ (7.03 MeV) excited states of $^{14}$N. The calculations are performed 
in the super-parallel kinematics and with the SF-B overlap functions. 
The shape of the recoil-momentum distribution for each state is determined by 
the CM orbital angular momentum $L$ of the knocked-out pair. Different partial 
waves of relative and CM motion contribute to the TOF. Each transition is
characterized by different components, with specific values of $L$. The 
relative weights of these components, which are given by the two-nucleon 
removal amplitudes included in the TOF, determine the shape of the 
recoil-momentum distribution \cite{barb}. 

The CM contribution included  in the orthogonalized approach gives in 
fig.~\ref{pncm3} an enhancement of the OB cross section that depends on the 
final state considered. The results are shown in the right panels of the figure. 
For the ground state the enhancement is of about the same size as the one found
in fig.~\ref{pncm1} for the $1_2^+$ state. A smaller effect is obtained for the 
$0^+$ and $2^+$ states. The enhancement is large for all the states considered 
in the figure at low values of  $p_B$ and is strongly reduced beyond 
200 MeV/$c$,  where for the $0^+$ and $2^+$ states the difference between the 
results of the two approaches becomes very small. 
The difference is reduced in the left panels, where also the contribution of 
the TB currents is included. For the $1_1^+$ ground state a significant 
enhancement of the final cross section is obtained in the orthogonalized model 
at low values of $p_B$. The sensitivity of the results to the parametrization 
used for the TB currents is for this state similar to the one found for the  
$1_2^+$ state in fig.~\ref{pncm1}. For the 
$0^+$ and $2^+$ states, where the TB currents play a more important role, the 
sensitivity of the calculated cross sections to the parameters used in the 
TB currents (dashed vs.\ solid) 
 is somewhat larger and generally larger than the difference 
produced by the use of orthogonal s.p. wave functions
 (dotted vs.\ dashed). The effects due to the orthogonalization 
are very small and even negligible at high values of $p_B$. It can be noted 
that when also the TB currents are included, the final cross section to the 
$0^+$ state calculated in the orthogonalized approach is lower than the one
calculated with our previous approach \cite{barb}.
This result is due to the different and combined effect of the different 
components of the TB current. 
The cross section calculated with the Bonn parametrization are  generally lower
than the ones calculated with the unregularized parametrization. 

In the $^{16}$O(e,e$'$pn)$^{14}$N measurements reported in \cite{middleton} the
energy resolution was sufficient to distinguish groups of states in the residual
nucleus but not good enough to separate individual states. Therefore, the cross
sections were measured for the group of states in the excitation energy range
between 2 and 9 MeV and include the $0^+$, $1_2^+$, and $2^+$ states. 
The present results in figs.~\ref{pncm1} and \ref{pncm3} indicate that the 
$1_2^+$ state dominates the cross section, the contribution of the $2^+$ state 
is competitive only for recoil-momentum values above 250 MeV/$c$, and the
contribution of the $0^+$ state is always negligible. 
The strong enhancement of the cross section to the $1_2^+$ state at low values
of $p_B$, that is due to the CM effects in the OB current included in the 
orthogonalized approach, is able to resolve the discrepancies found 
in \cite{middleton} in comparison with the experimental data and give a much 
better agreement \cite{secondpn}. A careful comparison with the data  requires 
calculations for a number of kinematical settings covering the energy and 
angular ranges subtended by the experimental set-up and will be presented 
in a forthcoming paper \cite{secondpn}.

\begin{figure}
\centerline{
\resizebox{0.55\textwidth}{!}{
  \includegraphics{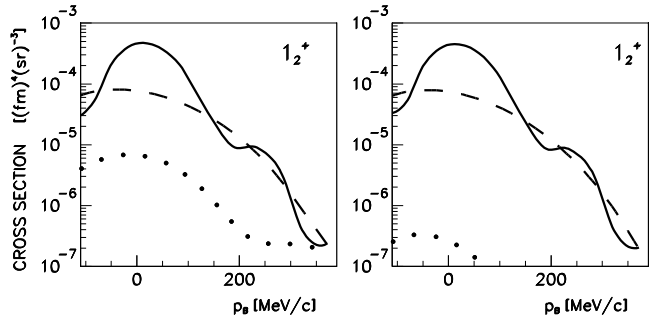}}
}
\caption{
The differential cross section of the $^{16}$O(e,e$'$pn)$^{14}$N reaction to 
the $1_2^+$ state as a function of $p_B$ in the 
same super-parallel kinematics as in fig.~\ref{pncm1}.
The final results given by sum of the OB and the TB currents 
are displayed in the left panel, the separate contribution of the OB  
current is shown in the right panel. The calculations are performed in the 
orthogonalized approach and with the Bonn parametrization for the TB
currents. Results obtained with different TOF's are compared: SF-B (solid), 
SF-CC (dashed), and SM-SRC (dotted). 
}
\label{pncm4}
\end{figure}
At next, we discuss the central point of these studies, namely 
 the sensitivity of the cross sections to  the treatment of correlations in 
the TOF. The cross sections calculated with the TOF's of \cite{barb} 
(SF-B), \cite{GMPS} (SF-CC), and \cite{GPpn} (SM-SRC) are compared  in 
fig.~\ref{pncm4}. The calculations are performed with the orthogonalized 
approach for the reaction $^{16}$O(e,e$'$pn)$^{14}$N to the $1_2^+$ state in 
the super-parallel kinematics. The three overlap functions give large 
differences, both on the shape and the magnitude of the calculated cross 
sections.
With the simpler SM-SRC overlap function, where only SRC are taken into account,
the contribution of the OB current is negligible and is up to about three 
orders of magnitude lower than the one obtained with the more complete and 
sophisticated approach SF-B. When the TB currents are added, the SM-SRC cross
section is enhanced by about one order of magnitude, the difference between
the SM-SRC and SF-B cross sections is reduced but remains very large, up to 
about two orders of magnitude in the maximum region and somewhat smaller at 
high values of $p_B$. 
The results in fig.~\ref{pncm4} have been obtained with the Bonn 
parametrization for the TB currents. A calculation with the parametrization used
in our previous calculations enhances the contribution of the $\Delta$-current,
but does not change significantly the final cross sections and therefore the 
main features of the results shown in the figure.   

The SM-SRC cross section is dominated by the TB currents and, as such, it is
practically unaffected by the use of orthogonalized s.p. bound and scattering
wave function. Incidentally, we note that for the separate contribution of 
the OB current the CM effects included in the orthogonalized approach 
give with SM-SRC a much smaller effect than with SF-B and SF-CC. 
This result confirms what was already obtained in \cite{cm}, {\it i.e.} that the
relevance of these CM effects depends on the TOF used in the calculation.  
The very large enhancement of the OB current contribution due to these CM 
effects makes the OB current dominant in the cross sections calculated with 
the SF-B and SF-CC overlap functions. We note that in the calculations 
of \cite{barb} and \cite{GMPS} the cross section to the $1_2^+$ state obtained
with SF-B and SF-CC was in both cases dominated by the TB $\Delta$-current. 

The much larger contribution of the OB current with the SF-B and SF-CC overlap 
functions emphasizes the crucial role played by TC, that are very important in 
proton-neutron emission and are neglected in the simpler SM-SRC
calculation. In the SF-B and SF-CC overlap functions SRC and TC are accounted
for consistently in the defect functions, which are calculated in the two 
TOF's within different methods. 

In fig.~\ref{pncm4} the SF-B cross section is generally larger than the 
SF-CC one.  The SF-B result overshoots the SF-CC one up to a factor of 6 in the 
maximum region. The differences are strongly reduced for values of
$p_B$ greater than 100 MeV/$c$. 

The differences between the cross sections calculated with SF-B and SF-CC are
due to the different treatments of all the contributions  to the TOF.  
A more complete calculation of LRC in an extended shell-model basis is 
performed in SF-B \cite{barb}. Moreover, the normalization of the two-nucleon 
overlap amplitudes is higher in the SF-B calculation. The difference in the 
shape of the cross section is due to the different mixing of configurations in 
the two cases. The defect functions are not only mixed differently by the 
different two-nucleon removal amplitudes, but different models are used to 
generate them in the two calculations, as well as different NN-interactions: 
Bonn-C in \cite{barb} Argonne V14 in \cite{GMPS}.  

The effect of the mutual interaction between the two outgoing nucleons (NN-FSI) 
has been neglected in the calculations presented till now. 
NN-FSI has been studied within a perturbative treatment in \cite{sch1a,sch1}, 
where it is found that the effect depends on the kinematics, on the type of 
reaction, and on the final state of the residual nucleus. 
NN-FSI effects are in general larger in pp- than in pn-knockout.  
For the $^{16}$O(e,e$'$pn)$^{14}$N reaction in the super-parallel
kinematics the effects of NN-FSI were found small but non negligible
  \cite{sch1}. 
The calculations in \cite{sch1} were performed with SF-CC
\cite{GMPS}. Since NN-FSI is sensitive to the various ingredients of the
calculations, it can be interesting, also in view of the comparison with the 
(e,e$'$pn) data of \cite{middleton}, to investigate its effects in the present 
orthogonalized  approach.  

\begin{figure}
\centerline{
\resizebox{0.55\textwidth}{!}{
  \includegraphics{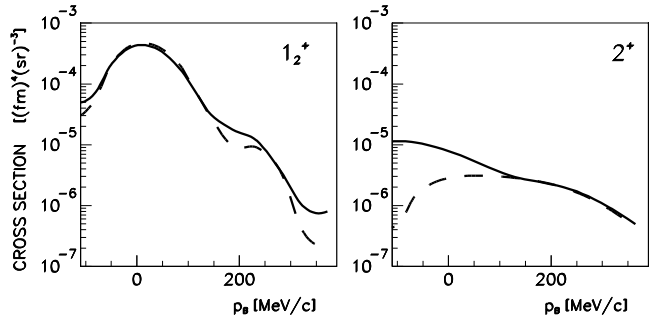}}
}
\caption{
The differential cross section of the $^{16}$O(e,e$'$pn)$^{14}$N reaction to 
the $1_2^+$ (3.95 MeV) and $2^+$ (7.03 MeV) excited states of $^{14}$N as a 
function of $p_B$ in the same super-parallel kinematics as in fig.~\ref{pncm1}.
The calculations are performed in the orthogonalized approach and with the Bonn 
parametrization for the TB currents. TOF as in fig.~\ref{pncm1}. 
Line convention: DW-NN (solid), DW (dashed).
}
\label{pncm5}
\end{figure}
The contribution of NN-FSI to the cross section of the reaction 
$^{16}$O(e,e$'$pn)$^{14}$N to the $1_2^+$ and $2^+$ excited states of 
$^{14}$N in the super-parallel kinematics is shown in  fig.~\ref{pncm5}. 
The results obtained in the approach considered till now (DW), where 
only the interaction of each one of the outgoing nucleons with the residual 
nucleus is considered, are compared with the results of the more complete 
treatment (DW-NN) where also the mutual interaction between the two outgoing 
nucleons is included within the same perturbative approach as in \cite{sch1}. 
The calculations have been performed in the orthogonalized approach with the 
more refined SF-B overlap function and with the Bonn parametrization for the TB
currents. 
For the $1_2^+$ state the contribution of NN-FSI is quite moderate.
It is practically negligible at low values of $p_B$, where the cross section has
its maximum, and it is appreciable at high missing momenta, where the cross 
section is much lower. In this region the slight enhancement and change of 
shape produced by NN-FSI might be appreciated in the comparison with the 
 data \cite{secondpn}. 
For the $2^+$ state NN-FSI enhances the cross section up to about one order of
magnitude at low values of $p_B$, where, however, the cross 
section remains completely dominated by the $1_2^+$ state, and is negligible 
above 100 MeV/$c$, and, therefore, also for the higher recoil momenta where the
contribution of the  $2^+$ state can be comparable to the one of the 
$1_2^+$ state.

\begin{figure}
\centerline{
\resizebox{0.55\textwidth}{!}{
  \includegraphics{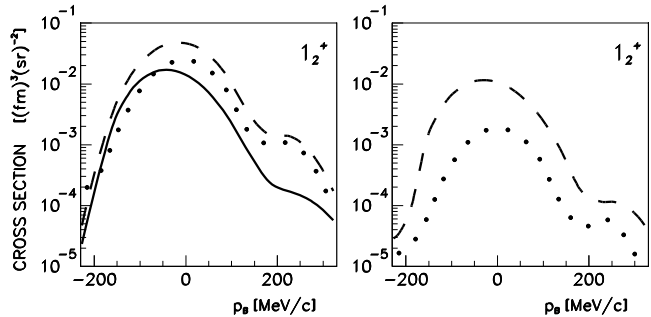}}
}
\caption{
The differential cross section of the $^{16}$O($\gamma$,pn)$^{14}$N reaction 
to the $1_2^+$ state as a function of $p_B$ in a 
super-parallel kinematics with $E_\gamma=215$ MeV. Calculations are performed in
the DW approach and with the same TOF as in fig.~\ref{pncm1}. 
Line convention for left and right panels as in fig.~\ref{pncm1}.
}
\label{pncm6}
\end{figure}

The comparison between the cross sections obtained in the present approach, 
where orthogonality is enforced between s.p. bound and scattering states, 
and in the previous approach, where the contributions of the OB current without 
correlations is subtracted in the transition amplitude, is presented in 
fig.~\ref{pncm6} for the photoinduced reaction 
 $^{16}$O($\gamma$,pn)$^{14}$N to the $1_2^+$  
excited state of $^{14}$N. Calculations have been performed in a super-parallel 
kinematics and for an incident photon energy which has the same value,  
$E_\gamma=215$ MeV, as the the energy transfer in the (e,e$'$pn) calculations 
of  figs.~\ref{pncm1}-\ref{pncm5}. Although the super-parallel kinematics is 
not well suited for ($\gamma$,pn) experiments, this case can be interesting 
for a theoretical comparison with the corresponding results of the electron 
induced reaction.    

The orthogonalized approach enhances the cross section calculated with the OB
current. This effect, that is shown in the right panel of  fig.~\ref{pncm6}, is
large although a bit lower than the one found in the corresponding situation for
the (e,e$'$pn) reaction in fig.~\ref{pncm1}. The difference between the two
results is significantly reduced when also the contribution of the TB currents 
is included. The final cross sections calculated in the two approaches differ 
at most by a factor of about 2
 (dotted vs.\ dashed).  As becomes obvious from the left panel, 
 a larger difference is given in the case of the 
($\gamma$,pn) reaction by the treatment of the TB currents. 
The cross section calculated with the orthogonalized approach and with 
the Bonn parametrization (solid) is 
 lower than the one obtained with the previous 
approach and the unregularized parametrization (dotted). 

\begin{figure}
\centerline{
\resizebox{0.55\textwidth}{!}{
  \includegraphics{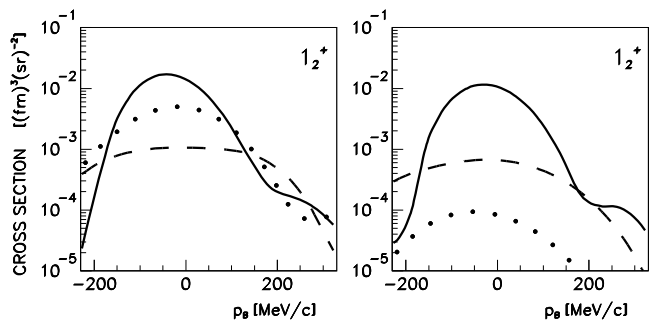}}
}
\caption{
The differential cross section of the $^{16}$O($\gamma$,pn)$^{14}$N reaction 
to the $1_2^+$ state as a function of $p_B$ in the same  
super-parallel kinematics as in fig.~\ref{pncm6}. Line convention for left and 
right panels as in fig.~\ref{pncm4}.
}
\label{pncm7}
\end{figure}

The cross sections displayed in fig.~\ref{pncm6} have been calculated with the
SF-B overlap function. The results obtained, with the orthogonalized approach 
and the Bonn parametrization, for the different TOF's are compared in 
fig.~\ref{pncm7}.
Also in this case large differences are found in the shape and in the magnitude
of the calculated cross sections. The contribution of the OB current calculated
with SF-B and SF-CC is much larger than with SM-SRC. The difference is,
however, less dramatic than in the corresponding situation of fig.~\ref{pncm4} 
for the (e,e$'$pn) reaction. In the final cross sections calculated with SF-B
and SF-CC  the TB currents produce a moderate enhancement of the OB 
contribution. In contrast, with SM-SRC the cross section is dominated by the 
TB currents which enhance the OB cross section by more than one order of 
magnitude. As a consequence of this enhancement, the final result with SM-SRC 
turns out to be closer to the SF-B one.

\begin{figure}
\centerline{
\resizebox{0.35\textwidth}{!}{
  \includegraphics{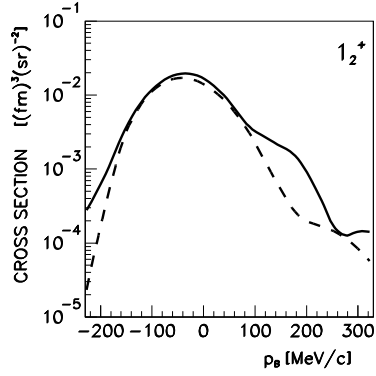}}
}
\caption{
The differential cross section of the $^{16}$O($\gamma$,pn)$^{14}$N reaction 
to the $1_2^+$ state as a function of $p_B$ in the same  
super-parallel kinematics as in fig.~\ref{pncm6}. The calculations are 
performed in the orthogonalized approach and with the Bonn parametrization for 
the TB currents. TOF as in fig.~\ref{pncm1}. Line convention: DW-NN (solid), 
DW (dashed)
}
\label{pncm8}
\end{figure}
The effect of NN-FSI on the cross section of the reaction 
$^{16}$O($\gamma$,pn)$^{14}$N to the $1_2^+$ state in the 
super-parallel kinematics is shown in  fig.~\ref{pncm8}. NN-FSI enhances the
cross section. This effect is very small for low values of $p_B$ and increases 
at higher values, where the cross section decreases.

\begin{figure}
\centerline{
\resizebox{0.55\textwidth}{!}{
  \includegraphics{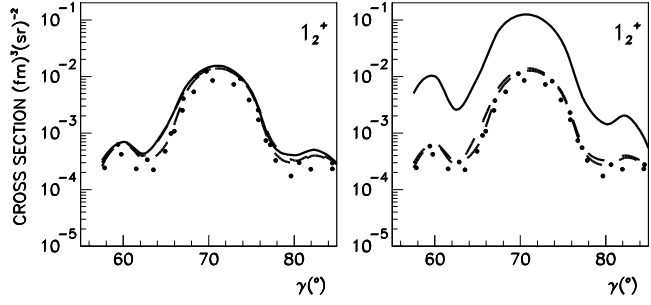}}
}
\caption{
The differential cross section of the 
$^{16}$O($\gamma$,pn)$^{14}$N  reaction to the $1_2^+$  state in a coplanar 
symmetrical kinematics with $E_\gamma=400$ MeV as a
function of the scattering angle of the outgoing nucleons.
Calculations are performed in the orthogonalized approach and with the same 
TOF as in fig.\ref{pncm1}, i.e.\ SF-B.
  The dotted lines give the separate contribution of 
the OB current, the dotted-dashed lines the sum of the OB and 
seagull currents, the dashed lines the sum of the OB, seagull, and 
pion-in-flight currents, and the solid lines the final result, where also the 
contribution of the $\Delta$-current is included. 
The Bonn parametrization is used for the TB currents in the left panel, the
unregularized parametrization in the right panel. FSI are treated within the DW 
approach.
}
\label{pncm9}
\end{figure}
A different kinematical situation is considered in fig.~ \ref{pncm9}, where the  
cross section of the reaction $^{16}$O($\gamma$,pn)$^{14}$N to the $1_2^+$ 
state has been calculated with an incident photon energy 
$E_\gamma=400$ MeV in a coplanar symmetrical kinematics, where the two nucleons 
are ejected at equal energies and equal but opposite angles with respect to 
the momentum transfer. 
In this kinematical setting different values of $p_B$ are obtained 
changing the scattering angles of the two outgoing nucleons.
In this case the cross sections are practically unaffected by the use of
orthogonalized s.p. bound and scattering wave functions and also the effect of
NN-FSI is small. The results are sensitive to the TB currents and to their
treatment.  The cross sections calculated with the SF-B overlap
function for the two parametrizations are compared in the left and right 
panels. With the Bonn parametrization the enhancement produced by the TB
currents in the cross section is within a factor of 2 and the main contribution
to this enhancement is given by the seagull current. A strong enhancement of 
the $\Delta$-current contribution is produced by the unregularized
 parametrization. In this case the contribution of the
$\Delta$-current is dominant and increases the cross section by more than one 
order of magnitude.

\begin{figure}
\centerline{
\resizebox{0.55\textwidth}{!}{
  \includegraphics{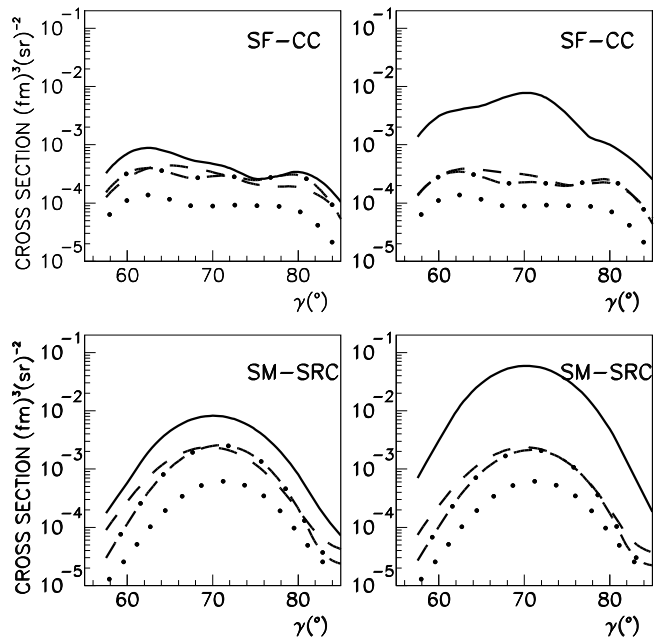}}
}
\caption{
The differential cross section of the 
$^{16}$O($\gamma$,pn)$^{14}$N  reaction to the $1_2^+$  state as a function of 
the scattering angle of the outgoing nucleons  in the 
same coplanar symmetrical kinematics as in fig.~\ref{pncm9}.
Results obtained with different TOF's are displayed:  SF-CC (top panels), 
SM-SRC (bottom panels).
Line convention for right and left panels as in fig.~\ref{pncm9}.
}
\label{pncm10}
\end{figure}
The corresponding cross sections calculated with the SF-CC and SM-SRC overlap 
functions for the reaction $^{16}$O($\gamma$,pn)$^{14}$N to the $1_2^+$ state 
in the coplanar symmetrical kinematics with $E_\gamma=400$ MeV  
are displayed in fig.~\ref{pncm10}. With both SF-CC and SM-SRC, as well as with
SF-B in   fig.~\ref{pncm9}, the cross section is sensitive to the
parametrization used for the TB currents. When the unregularized 
parametrization is used in the calculations, the $\Delta$-current is dominant 
both with SM-SRC and SF-CC. With SM-SRC the $\Delta$-current gives 
the main contribution also when the Bonn parametrization is used. 
For every TOF here considered the cross sections obtained with the two 
pa\-ra\-met\-rizations differ by about one order of magnitude, while the shape is not 
particularly affected by the treatment of the TB currents. 
In contrast, large differences, both in the shape and in the magnitude of the 
calculated cross sections, are obtained with different TOF's. 
This confirms that the treatment of correlations in the two-nucleon wave 
function affects all the ingredients of the calculations, and not only the 
  OB current but also the TB currents.

\section{Summary and conclusions}

\label{sec:conclusions}

Electromagnetically induced two-nucleon  knockout is an ideal tool to
 study the role of  correlations in the nuclear wave function
 going beyond the shell-model approach. In this paper, we have 
presented recent improvements in the theoretical model of proton-neutron 
knockout, which is of specific interest for the study of tensor correlations, 
which are suppressed in  proton-proton knockout.  

In comparison with earlier studies, a more complete treatment of CM effects 
has been included in the model. In the CM frame the transition operator 
becomes a two-body operator even in the case of a one-body nuclear current. 
As a consequence, the one-body current can give a contribution to the 
cross section of two-particle emission independently of correlations. These 
CM effects were not properly taken into account in our previous 
calculations for pn-knockout. They have been included in this work enforcing 
orthogonality between s.p. initial and final states by means of a Gram-Schmidt 
orthogonalization. This procedure, that was applied in \cite{cm} to pp-knockout,
allows us to naturally include all the CM effects as well as to get rid of 
possible spurious contributions to two-nucleon emission, due to the lack of 
orthogonality between bound and scattering states obtained by the use of an 
energy-dependent optical model potential.
 
The treatment of the two body seagull, pion-in-flight and 
$\Delta$-currents has been improved using a more realistic regularized 
approach which is consistent with the description of elastic NN-scattering 
data. 
The role of the mutual interaction between the two outgoing nucleons, that 
is usually ignored,  has been reconsidered  in combination with the present 
improvements.
 Last but not least, the  sensitivity of the cross sections to NN-correlations
 has been studied comparing results obtained with different 
 two-nucleon overlap functions, where correlations are included with more 
 refined or simpler approaches. 

 It turns out that the effect of these different aspects strongly depends
 on the chosen kinematics. In the super-parallel kinematics, which is of
 particular interest with respect to experiment, the CM effects included with 
 the enforced orthogonalization lead to a dramatic increase, up to an order of
 magnitude, of the contribution of the one-body current, which becomes dominant 
 in the $^{16}$O(e,e$'$pn)$^{14}$N reaction. In the final cross section the 
 enhancement is particularly large for the transition to the $1_2^+$ 
 (3.95 MeV) excited state of $^{14}$N and for low values of the missing 
 momentum. 
 For different final states or in different kinematics the influence of these 
 effects is small or even negligible. In particular, it is negligible when the 
 cross section is dominated by the two-body currents, that are not sensibly 
 affected by the orthogonalization procedure.
   
 The regularized treatment of the two-body currents leads in general to a 
 dramatic reduction of the $\Delta$-current contribution  with respect to the 
 results obtained with the previous treatment. 
 In contrast, the corresponding effects on the nonresonant seagull and 
 pion-in-flight  meson-exchange currents are of minor importance. 
The difference due to the regularized and unregularized parametrizations on 
the final cross section depends on the role played by the $\Delta$-current. 
 In the super-parallel kinematics for the (e,e$'$pn) reaction, the latter is
only of minor importance so that  
 the resulting difference between the different parametrizations 
is within a factor of about 2. Larger differences are 
obtained in the  ($\gamma$,pn) reaction, where in the symmetrical kinematics 
the regularized parametrization reduces the cross section by more than one 
order of magnitude.

The effect of the mutual interaction between the two outgoing nucleons is in
general moderate although not negligible. This contribution does not change 
the qualitative features of the cross sections but it should be evaluated for 
a more careful comparison with the experimental data.
  
Dramatic differences, both in the shape and magnitude of the calculated cross 
sections are given, in all the considered situations and kinematics, by 
different treatments of correlations in the two-nucleon wave function. 
Correlations affect both one-body and two-body current contributions. 
A crucial role is played by tensor correlations. When tensor correlations are 
neglected in the overlap function the contribution of the one-body current to 
the cross section is strongly underestimated and becomes always negligible.  
The cross sections of the $^{16}$O(e,e$'$pn)$^{14}$N reaction in super-parallel 
kinematics differ up to about two orders of magnitude depending on whether 
tensor correlations are taken into account or not.

In addition, also  a careful treatment of long-range correlations appears
 to be essential. In the calculations long-range correlations affect the 
 two-nucleon removal amplitudes. These amplitudes determine the weights of the 
 partial waves of relative and CM motion in the overlap function 
 and may therefore affect the shape and also the magnitude of the cross section. 
 
 In conclusion, we may expect that the comparison with presently available 
 data may lead to important conclusions about the structure of correlations 
 and therefore to a test of our present understanding of nuclear 
 structure in general. This topic will be outlined in future
  work \cite{secondpn}.

\section*{}
We thank Peter Grabmayr and Duncan Middleton for useful discussions. We 
are grateful to Carlo Barbieri and Herbert  M\"uther for providing us with 
the two-nucleon overlap functions that have been used in the calculations.



\end{document}